\documentclass[journal,10pt]{IEEEtran}
\usepackage[final]{graphicx}
\usepackage{multicol}
\graphicspath{{Images//}} 

\usepackage{soul,color}
\usepackage{psfrag}
\usepackage{epstopdf}
\usepackage{amsfonts}
\usepackage{mathrsfs}
\usepackage{pifont}
\usepackage{verbatim}
\usepackage{upgreek}
\usepackage{color}
\usepackage[usenames,dvipsnames]{xcolor}
\usepackage{caption}
\usepackage{algorithm}
\usepackage{mdwmath}
\usepackage{mdwtab}
\usepackage{amssymb,amsmath}
\usepackage{amsmath}
\usepackage{amsthm}
\usepackage{epstopdf}
\usepackage{cite}
\usepackage{tikz}
\usepackage{scalefnt}
\usepackage{subfigure}
\usepackage{MnSymbol}
\usepackage{setspace}
\usepackage{psfrag}

\usepackage[font=scriptsize]{caption}
\usepackage{algpseudocode}

\algrenewcommand\algorithmicforall{\textbf{foreach}}
\algrenewcommand\algorithmicindent{.8em}

\label{newcommand}
\label{English Chars}
\newcommand{\bA}{\textbf{A}}

\newcommand{\bb}{\textbf{b}}
\newcommand{\bC}{\textbf{C}}

\newcommand{\bx}{\textbf{x}}

\newcommand{\by}{\textbf{y}}

\newcommand{\dom}{\mbox{\bf dom}\, }

\label{Roman Chars}

\label{Theorems}

\theoremstyle{remark}

\begin{document}
% Title
\label{title}
\title{Submodularity in Action: From Machine Learning to Signal Processing Applications}

\author{%
	\IEEEauthorblockN{Ehsan Tohidi\IEEEauthorrefmark{1},
		Rouhollah Amiri\IEEEauthorrefmark{2},
		Mario Coutino\IEEEauthorrefmark{3},
		David Gesbert\IEEEauthorrefmark{1},\\
		Geert Leus\IEEEauthorrefmark{3}, and
		Amin Karbasi\IEEEauthorrefmark{4}}\\
	\IEEEauthorblockA{\IEEEauthorrefmark{1}%
		Communication Systems Department, EURECOM,
		Biot, France
		}\\
	\IEEEauthorblockA{\IEEEauthorrefmark{2}%
		Electrical Engineering Department, Sharif University of Technology,
		Tehran, Iran
		}\\
	\IEEEauthorblockA{\IEEEauthorrefmark{3}%
		Faculty of Electrical Engineering, Mathematics and Computer Science, Delft University of Technology, Delft, The Netherlands %, 
		}\\
		\IEEEauthorblockA{\IEEEauthorrefmark{4}%
		Electrical Engineering and Computer Science Department, Yale University,
New Haven, USA
		}
}

\maketitle

\begin{abstract}
Submodularity is a discrete domain functional property that can be interpreted as mimicking the role of the well-known convexity/concavity properties in the continuous domain. Submodular functions exhibit strong structure that lead to efficient optimization algorithms with provable near-optimality guarantees. These characteristics, namely, efficiency and provable performance bounds, are of particular interest for signal processing (SP) and machine learning (ML) practitioners as a variety of discrete optimization problems are encountered in a wide range of applications. Conventionally, two general approaches exist to solve discrete problems: $(i)$ relaxation into the continuous domain to obtain an approximate solution, or $(ii)$ development of a tailored algorithm that applies directly in the discrete domain. In both approaches, worst-case performance guarantees are often hard to establish. Furthermore, they are often complex, thus not practical for large-scale problems. In this paper, we show how certain scenarios lend themselves to exploiting submodularity so as to construct scalable solutions with provable worst-case performance guarantees. We introduce a variety of submodular-friendly applications, and elucidate the relation of submodularity to convexity and concavity which enables efficient optimization. With a mixture of theory and practice, we present different flavors of submodularity accompanying illustrative real-world case studies from modern SP and ML. In all cases, optimization algorithms are presented, along with hints on how optimality guarantees can be established.
\end{abstract}

%% ***************** Signal Model **********************
\section{Introduction}
\subsection{Discrete Optimization}
Discrete optimization is a notoriously challenging problem which occurs in countless engineering applications and particularly, in SP and ML. Discrete optimization usually involves finding a solution in some finite or countably infinite set of potential solutions that maximizes/minimizes an objective function. A counter-intuitive phenomenon is that discrete problems are sometimes more difficult than their continuous counterparts. This phenomenon was perfectly illustrated by Welsh \cite{welsh1976matroid}: ``mathematical generalization often lays bare the important bits of information about the problem at hand".

In general, discrete optimization problems are tackled in two common ways: $(i)$ building a continuous relaxation or $(ii)$ working out a tailored algorithm. The first relaxes the original problem to a continuous one so as to apply tools from continuous optimization. The continuous solution is then made discrete by a rounding technique to obtain an approximate feasible solution for the original problem. The second approach is to develop a customized algorithm to directly use in the discrete domain. Beyond the fact that we often need creative endeavor to design these algorithms, both approaches have the following shortcomings: $(i)$ for most of the problems, it is hard to have a gap bound between the approximate and optimal solutions, i.e., approximation guarantees; and $(ii)$ although the proposed algorithms are usually solvable in polynomial time, they are not necessarily scalable.

\subsection{Submodularity: A Useful Property for Optimization}
In this paper, we depict a family of optimization scenarios and seek solutions that circumvent the above fundamental limitations. While it is challenging to find optimality bounds {\em and} provide low complexity methods that will address {\em all} discrete optimization scenarios, in this paper we highlight a particular structure that is relevant to a surprisingly large class of problems. 
Submodularity is a functional property which has recently gained significant attention.
The submodularity property enables striking algorithm-friendly features and is observed in a good number of application scenarios. 
In fact, due to specific connections with convexity, exact minimization of submodular functions can be done efficiently, while greedy algorithms can be shown to obtain near-optimality guarantees in submodular maximization problems thanks to relations with concavity.
These benefits have drawn attention in many different contexts 
~\cite{krause2014submodular,bach2013learning,fujishige2005submodular}. Sample applications include sensor selection \cite{8537943}, detection \cite{coutino2018submodular}, resource allocation \cite{7208844}, active learning \cite{golovin2011adaptive},
interpretability of neural networks \cite{elenberg2017streaming}, and adversarial attacks \cite{8263844}, to name a few.

This paper takes an illustrative approach to explain the concept of submodularity and its promise in modern SP and ML applications.
The relation of submodularity with the celebrated properties of convexity/concavity is introduced to motivate the algorithmic landscape for optimizing submodular functions.
Starting with basic structures, we give a presentation of different aspects of submodular functions and cover current extensions of submodularity.
Also, to clarify the importance of submodularity in the context of SP and ML, various applications are showcased as examples and 
their connections to different aspects of submodularity are highlighted. 
 
\section{Submodularity Essentials}
In this section, we explain the mathematical formulation of submodularity through an illustrative example.

\subsection{Motivating Example: Sensing Coverage Problem}
\begin{figure}
	\centering
	\includegraphics[width=.45\textwidth]{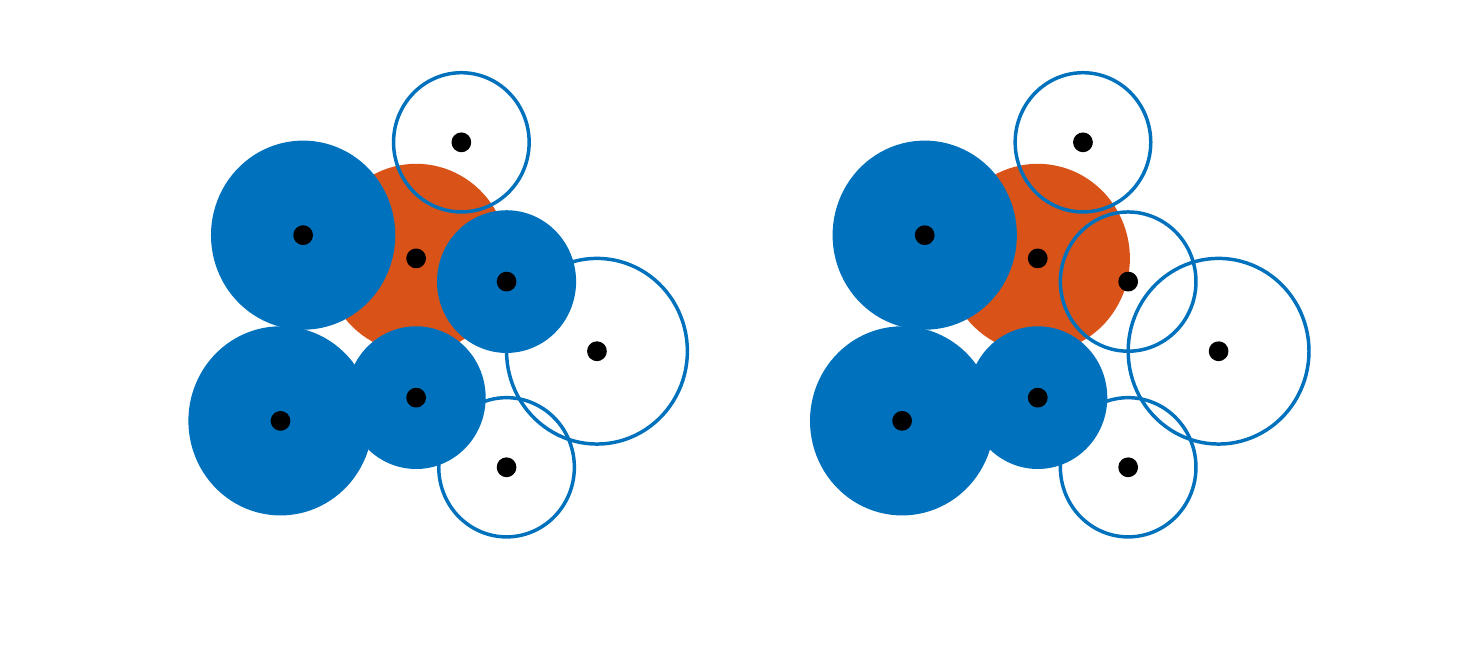}		
	\caption{Illustration of submodularity for the sensing coverage problem. The figure depicts an example of adding the same sensor (i.e., the red circle) to a large subset (left subfigure) and a smaller subset (right subfigure), where the selected subsets of sensors are shown by the blue circles. The amount of added coverage is larger in the right subfigure.}
	\label{warmup_example}
\end{figure}

We illustrate the concept of submodularity through the so-called sensing coverage problem. Assume there exists a set of candidate locations of sensors--called the ground set--with effective areas that determine the area around each sensor from where fruitful field measurements can be obtained (see Fig. \ref{warmup_example}). 
Since these sensors are costly, we should optimize the deployment locations in order to maximize their effectiveness. Therefore, the optimization problem turns into a coverage problem where the objective function is to maximize the covered area of the sensing field subject to a limited number of selected sensors. Here, we investigate one important property of this objective function.

Fig. \ref{warmup_example} depicts two sample subsets of the ground set (i.e., the blue circles) where the sensors in the smaller subset are all included in the larger one. 
We have added the same sensor (i.e., the red circle) to both subsets. As shown in Fig. \ref{warmup_example}, the new covered area due to the newly added sensor is smaller for the larger subset. We often call this phenomenon \textit{diminishing returns} or \textit{decreasing gain}. For this specific example, this property implies that the new covered area added due to the addition of a new sensor may decrease when the size of the original subset increases. As will be seen next, this example carries the essential intuition behind submodularity.

\subsection{Submodular Set Functions}
A set function $f:2^{\mathcal{N}}\rightarrow \mathbb{R}$ defined over a ground set $\mathcal{N}$
is submodular if for every $\mathcal{A}\subseteq \mathcal{B} \subseteq \mathcal{N}$ and $e\in \mathcal{N}\,\backslash\, \mathcal{B}$ it holds that
\begin{equation}
f(e|\mathcal{A}) \geqslant f(e|\mathcal{B}),
\label{def1}
\end{equation}
where $f(e|\mathcal{S}) = f(\mathcal{S}\cup\{e\}) - f(\mathcal{S})$ is defined as the discrete derivative (or marginal gain) of the set function $f$ at $\mathcal{S}$ with respect to $e$. Although there are several equivalent definitions of submodularity, we stick to the definition in \eqref{def1} as it naturally displays the diminishing returns property of submodular set functions. Moreover, a function $f$ is supermodular if $-f$ is submodular, and a function is modular if it is both submodular and supermodular.

Besides submodularity, another common property of some set functions is monotonicity. A set function $f:2^{\mathcal{N}}\rightarrow \mathbb{R}$ is monotone if for every $\mathcal{A}\subseteq \mathcal{B} \subseteq \mathcal{N}$, $f(\mathcal{A}) \leqslant f(\mathcal{B})$. This property is exhibited by many commonly encountered set functions in practice. For instance in the earlier example, coverage is monotone in the number of sensors.

Note that while  there is a substantial body of literature which addresses submodular
functions  with non-monotone behavior \cite{feige2011maximizing}, in this paper, we mainly consider normalized monotone submodular functions for ease of exposition. Here, a normalized set function implies $f(\emptyset)=0$.

In the following sections, we present crucial properties linked with submodularity and extensions which find applications in both SP and ML.

\section{Diminishing Returns and Concavity: Maximization}
In this section, we focus on the link between submodularity and concavity which can be leveraged for designing algorithms for the maximization of submodular functions.

\begin{figure}
	\centering
	\includegraphics[width=.45\textwidth]{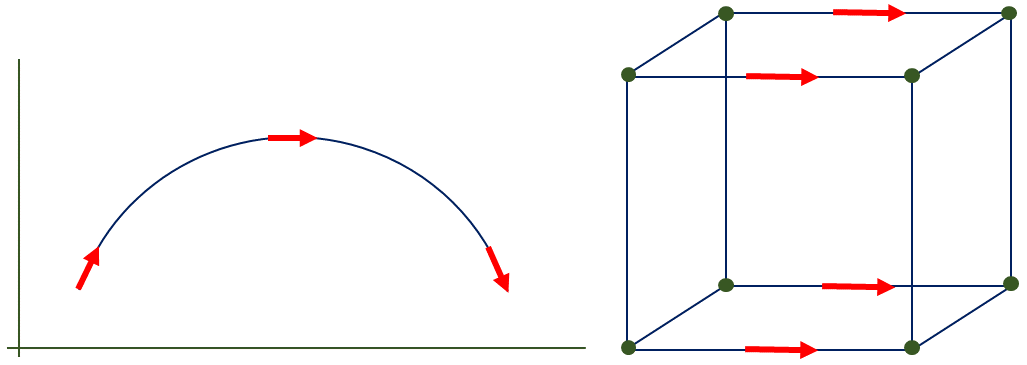}		
	\caption{Illustration of decreasing derivatives in both concave and submodular functions. The left and right subfigures depict a concave function in the continuous domain and a submodular function in the discrete domain, respectively.}
	\label{Concavity_illustrate}
\end{figure}

The definition \eqref{def1} exhibits the notion of diminishing marginal gains
which naturally leads to considering submodularity as a discrete analog of concavity. As depicted in the left subfigure of Fig. \ref{Concavity_illustrate}, the derivative of a concave function does not increase by increasing the input variable. Although less explicit, a similar characteristic appears in submodular functions. In the right subfigure of Fig. \ref{Concavity_illustrate}, a hypercube is shown where each vertex corresponds to a subset of the ground set and a line connects two vertices if one subset contains the other. Here, the arrows represent the direction from the smaller to the larger subset. Similar to the concavity in the continuous domain, moving along an arrow, the discrete derivative does not increase.

\subsection{Greedy Maximization of Submodular Functions}
Consider now the problem of submodular function maximization, which can be written in its most general form as
\begin{equation}
\begin{aligned}
\underset{\mathcal{S}\subseteq\mathcal{N}}{\max} \quad f(\mathcal{S}) \quad
\text{s.t.} \quad \text{some constraints on } \mathcal{S},
\end{aligned}
\label{equ:submaxgen}
\end{equation}
where $f(\cdot)$ is a submodular function and the common valid constraints are cardinality, knapsack, multi-partition, and matroid (more details are given later). Problem \eqref{equ:submaxgen} appears in many applications, for example, consider the previous sensing coverage problem [c.f. Fig.~\ref{warmup_example}]. In that setting, our goal is to deploy up to $K$ sensors, i.e., $|\mathcal{S}|\leqslant K$, such that the coverage area, i.e., $f(\mathcal{S})$, is maximized.
Finding the optimal solution of such a problem is NP-hard \cite{nemhauser1978analysis}. Therefore, it is reasonable to look for efficient algorithms that achieve a near-optimal solution.

\paragraph*{Greedy algorithm} A simple approach for solving the problem of normalized monotone submodular function maximization under a cardinality constraint is the greedy algorithm, which starts with an empty subset $\mathcal{S}$, and in each iteration adds the element $e$ that maximizes the marginal gain $f(e|\mathcal{S})$, i.e.,
\begin{align}\label{eq.greedy}
\mathcal{S}\leftarrow \mathcal{S}\cup\{\mathop{\arg\max } \limits_{e \in \mathcal{N}\setminus \mathcal{S}} \,f(e|\mathcal{S})\}.
\end{align}
This simple algorithm is guaranteed to achieve the optimal solution with a factor of $1-1/\text{e}$ \cite{nemhauser1978analysis}.
It is also known that one cannot achieve a better approximation guarantee under natural computational theoretic assumptions.
Although obtaining this performance guarantee with such a simple algorithm is surprising, yet there exists a good explanation. The definition \eqref{def1} is based on diminishing returns. 
Simply put, when we add elements to a set, the elements that are added in the beginning are more important than the elements that are included later. Now, a greedy algorithm only optimizes over the next element and does not consider later elements. As such, it gives the largest weight to the next element. This observation intuitively explains that the basis of a greedy algorithm is matched with the concept of submodularity and this is the main reason for achieving a noticeable performance by employing such an algorithm with a low computational complexity.

\subsection{Submodularity and Concavity}
A way to establish a connection between submodularity and concavity is achieved through the so-called multi-linear extension~\cite{vondrak2008optimal}.
For a set function $f: 2^{\mathcal{N}} \to \mathbb{R}$, its multi-linear extension $F_M: [0,1]^N \to \mathbb{R}$ with $|\mathcal{N}|=N$ is defined by
\begin{equation} \label{eq:ML Extension}
F_M(\bx) = \mathbb{E}\{f(\mathcal{X})\} = \sum_{\mathcal{S}\subseteq \mathcal{N}} f(\mathcal{S}) \prod_{i\in\mathcal{S}}x_i \prod_{j\in\mathcal{N}\setminus \mathcal{S}}(1-x_j),
\end{equation}
with $\mathcal{X}$ being a random set where elements appear independently with probabilities $x_i$.
It is proved that if $f(\cdot)$ is submodular, then $F_M(\cdot)$ is \emph{concave} along positive directions. Namely, if $\bx\leqslant \by$, then $\nabla F_M(\bx)\geqslant\nabla F_M(\by)$, where the inequality should be considered element-wise~\cite{vondrak2008optimal}, and $\nabla$ stands for the gradient operation.

Using the restricted concavity of the multi-linear extension, the continuous version of the greedy algorithm [cf.~\eqref{eq.greedy}], the so-called \emph{continuous greedy}, can be shown to solve certain types of constrained maximization problems near-optimally~\cite{calinescu2011maximizing}.

%%%%%%%%%%%%%%%%%%%%%%%%%%%%%%%%%%%%%%%%%%%%%%%%%%%%%%%
\section{Constrained Submodular Maximization}

In many applications, the elements of the ground set may have non-uniform costs, e.g., some sensors might be more expensive to deploy than others, and the problem may be constrained on a budget that the total cost cannot exceed or certain structure has to be enforced in the solution set. For example, consider a set of heterogeneous sensors (such as acoustic, optical, and radar), which provide a given sensing coverage and have different operative/deployment costs. We aim to deploy a number of these sensors to maximize the total coverage while meeting a budget requirement. Such a problem is called a \textit{knapsack problem} and appears often in SP and ML applications. This problem can be formulated as
\begin{equation}
\begin{aligned}
\underset{\mathcal{S}\subseteq\mathcal{N}}{\max} \quad f(\mathcal{S}) \quad
\text{s.t.} \quad \sum_{e\in\mathcal{S}}{c_e}\leqslant B,
\end{aligned}
\label{equ:submaxknap}
\end{equation}
where $c_e$ is the cost of element $e\in\mathcal{N}$ and $B$ is a given budget, determining the maximum sum-cost of elements in $\mathcal{S}$.

\paragraph*{Cost-Weighted Greedy Algorithm} The optimization problem \eqref{equ:submaxknap} is well-solved with a modified version of the greedy algorithm which takes the cost into account. This algorithm greedily generates a solution substituting the update selection rule~\eqref{eq.greedy} by
\begin{align}
\mathcal{S}\leftarrow \mathcal{S}\cup\{\underset{e\,\in\,\mathcal{J}}{\arg\max} \frac{f(e|\mathcal{S})}{c_e}\},
\end{align}
where $\mathcal{J}=\{e|e\in\mathcal{N}\,\backslash\,\mathcal{S}, c_e\leqslant B-c_{\mathcal{S}} \}$.
Next to this set, another greedy set is constructed using the rule~\eqref{eq.greedy}. It is shown that constructing both sets and selecting the best one, provides a $(\frac{1-1/\rm{e}}{2})$-approximation guarantee \cite{leskovec2007cost}. A later study showed that a more involved version of this procedure achieves a $(1-1/\sqrt{\rm{e}})$-approximation guarantee \cite{lin2010multi}. Further, if partial enumeration of all feasible sets of cardinality one or two is performed, a version of the cost-weighted greedy algorithm leads to a $(1-1/\rm{e})$-approximation guarantee \cite{sviridenko2004note}.

Although knapsack-type constraints are pervasive, they only model weights (or cost) associated to elements of the ground set. Fortunately, there exist other structures that are able to capture more complex and practical constraints while allowing for near-optimal maximization of submodular functions. In the following, we introduce them, along with their corresponding greedy algorithms.

\subsection{Matroids: Useful Combinatorial Structures}
The matroid is a useful combinatorial structure which generalizes the concept of linear independence in linear algebra to set theory.
Matroids and submodular functions are closely related. Specifically, each matroid corresponds to a submodular rank function.

A matroid is defined as a pair $(\mathcal{N}, \mathcal{I})$ in which $\mathcal{N}$ is a finite set and $\mathcal{I}\subseteq 2^{\mathcal{N}}$ comprises any subset of $\mathcal{N}$ which satisfies the following properties: 
\begin{itemize}
	\item $\mathcal{A}\subseteq \mathcal{B}\subseteq \mathcal{N}$ and $\mathcal{B}\in \mathcal{I}$ implies $\mathcal{A}\in \mathcal{I}$.
	\item $\mathcal{A},\mathcal{B}\in \mathcal{I}$ and $|\mathcal{B}|>|\mathcal{A}|$ implies that $\exists e \in \mathcal{B}\,\backslash\,\mathcal{A}$ such that $\mathcal{A}\cup\{e\}\in\mathcal{I}$.
\end{itemize}
Based on the above definition, the sets in $\mathcal{I}$ are called independent.
\paragraph*{Examples of matroids} In the following, we list some examples of commonly encountered matroids in signal processing and machine learning applications.
\begin{itemize}
    \item {Graphic matroid:} Consider a graph $\mathcal{G} = (\mathcal{V},\mathcal{E})$ with a vertex set $\mathcal{V}$ and an edge set $\mathcal{E}$ and let $\mathcal{I}(\mathcal{G})$ be the set of all edge subsets that do not contain a cycle of $\mathcal{G}$; that is, the elements of the independent sets are the forests of the graph. Then, $(\mathcal{E},\mathcal{I}(\mathcal{G}))$ forms a matroid. 
    \item {Uniform matroid:} Another common example of a matroid is given by a cardinality constraint, i.e., $|\mathcal{S}|\leq K$. The related matroid is constructed by considering as independent sets all the subsets of $\mathcal{N}$ with at most $k$ elements, i.e., $\mathcal{I} = \{\mathcal{S} \subset \mathcal{N} : |\mathcal{S}| \leq k\}$. This matroid is referred as the \emph{uniform matroid} of rank $k$, $\mathcal{U}_{\mathcal{N}}^k$. 
    \item {Partition matroid:} Given a collection of $I$ disjoint sets, $\{\mathcal{C}_i\subset\mathcal{N}\}_{i=1}^I$, and integers $\{b_i\}_{i=1}^I$ such that $0 \leq b_i \leq |\mathcal{C}_i|,\;\forall i\in\{1,\ldots,I\}$. Then, the independent sets of a \emph{partition matroid} are given by $\mathcal{I} = \{\mathcal{S}\subseteq\mathcal{N} : |\mathcal{S} \cap \mathcal{C}_i|\leq b_i,\; \forall i\; 1\leq i\leq I\}$.
\end{itemize}

\paragraph*{Matroid-Aware Greedy Algorithm} Given the matroid $(\mathcal{N}, \mathcal{I})$, the constrained submodular maximization problem $\mathop {\max}\limits_{\mathcal{S}\in \mathcal{I}}f(\mathcal{S})$ can be near-optimally solved by constructing a solution using the rule
\begin{align}
	\mathcal{S}\leftarrow \mathcal{S}\cup\{\mathop{\arg\max } \limits_{e\,\notin\,\mathcal{S}\,:\,\mathcal{S}\cup\{e\} \,\in\, \mathcal{I}} \,f(e|\mathcal{S})\},
\end{align}
until there is no more candidate element which can be added to form a feasible solution.
It is shown that the matroid-aware greedy method can achieve a $1/2$ near-optimality guarantee \cite{nemhauser1978analysis}.

The $(1-1/\text{e})$-approximation guarantee can be achieved if the \textit{continuous greedy} algorithm in \cite{calinescu2011maximizing} is used instead. This method constructs a solution by appropriately rounding the solution of a continuous relaxation, using the multi-linear extension, of the original problem.

In the following, we present an application that makes use of matroids to model commonly encountered constraints.

\subsection{Resource Selection for Parameter Estimation in MIMO Radars}

\begin{figure}[htb]
	%\vspace{-.3cm}
	\centering
	\centerline{\includegraphics[width=0.45\textwidth]{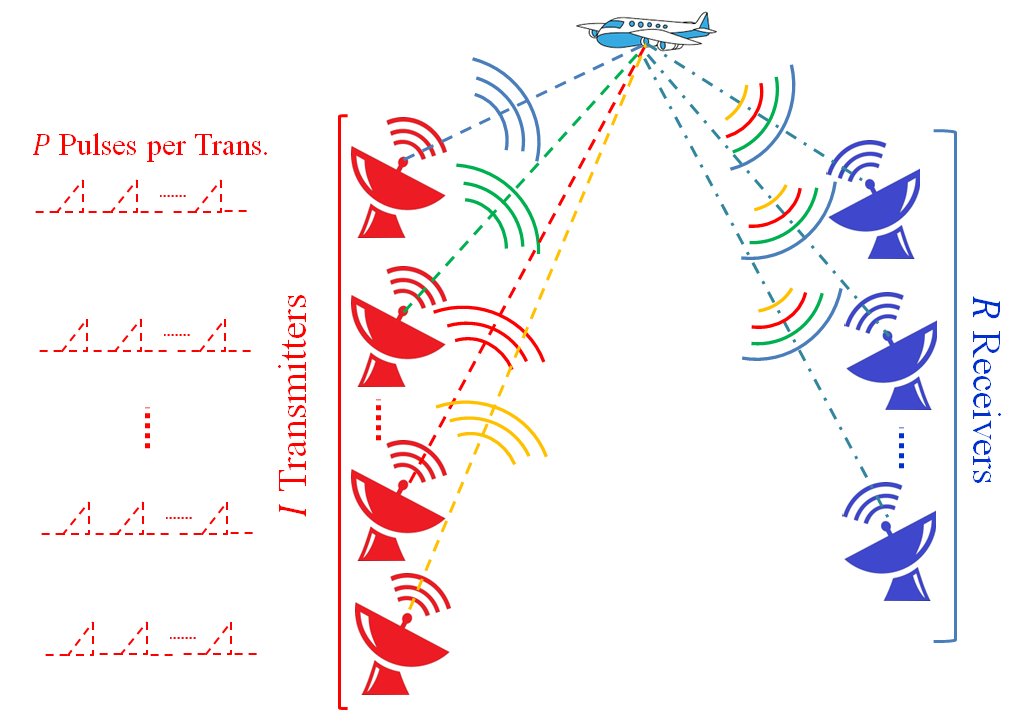}}
	%\vspace{-.2cm}
	\caption{ The configuration of a MIMO radar system. }
	\label{fig:MIMO}
	%\vspace{-.2cm}
\end{figure}

The application of multiple-input multiple-output (MIMO) radar systems becomes pervasive due to their enormous advantages over conventional radars.
Such large-scale MIMO systems are, however, very expensive to implement in practice, due to the high increase in hardware cost regarding the deployment of multiple sensors, the power consumption for multi-pulse transmissions, and the processing complexity.
To reduce the aforementioned costs and at the same time guaranteeing a given estimation accuracy level, it is meaningful to select only a limited set of transmitters, pulses, and receivers (shown in the MIMO radar configuration of Fig. \ref{fig:MIMO}) that are the most informative for the parameter estimation task. Such a problem is known as \textit{resource selection} in the literature.
In \cite{8537943}, the problem of resource selection in a MIMO radar is
formulated as maximizing a submodular function subject to a \textit{partition matroid} $(\mathcal{P}\cup\mathcal{R}, \mathcal{I})$ whose independent sets are defined as
\begin{align}
\mathcal{I} = \{\mathcal{S}:|\mathcal{S}\cap\mathcal{P}|\le K_P,|\mathcal{S}\cap\mathcal{R}|\le K_R\},
\end{align}
where $\mathcal{P}$ and $\mathcal{R}$ are the ground sets of all transmitted pulses and receivers, respectively. The variables $K_P$ and $K_R$ stand for cardinality of the selected sets of pulses and receivers.

\begin{figure}[htb]
	%\vspace{-.3cm}
	\centering
	\centerline{\includegraphics[width= 0.45\textwidth]{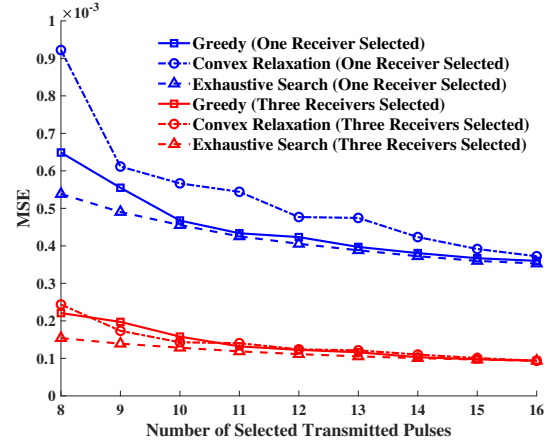}}
	%\vspace{-.2cm}
	\caption{ The MSE as a function of the selected transmitted pulses for two different cases with one and three selected receivers. Figure courtesy of \cite{8537943}. }
	\label{fig:MSE_Ant_Sel}
	%\vspace{-.2cm}
\end{figure}

To illustrate this application, we consider a simulation scenario with four receivers, four transmitters, and four pulses per transmitter.
To evaluate the performance of the greedy selection algorithm in comparison with the convex method and the optimum MSE obtained through an exhaustive search, the estimation MSE is plotted in Fig. \ref{fig:MSE_Ant_Sel} as a function of the number of selected transmitted pulses. The results are presented for two cases where one and three receivers should be selected. As shown in Fig. \ref{fig:MSE_Ant_Sel}, the estimation accuracy of the greedy algorithm is very close to the optimal value. Furthermore, its performance is the same or better than its convex counterpart, while having a much lower complexity.

The key to the problem is the objective function which should be related to the estimation accuracy. Since mean-square error (MSE) is neither convex nor submodular and makes the optimization task difficult, in \cite{8537943}, a surrogate objective function is incorporated which measures the orthogonality between vectors of a frame.
This measure has been shown to be an appropriate submodular proxy for the MSE in nonlinear estimation problems.

\subsection{Multi-way Partitions}

In addition to matroids, \emph{multiway-partitions} are amenable structures for submodular optimization. Multiway-partitioning arises in a diverse range of combinatorial optimization problems in areas including communications and signal processing. This problem is defined as partitioning a given set $\mathcal{S}$ into $k$ disjoint subsets $\mathcal{S}_1,\ldots,\mathcal{S}_k$, $\mathcal{S}_i\bigcap\mathcal{S}_j=\emptyset$, $\bigcup_{i=1}^k\mathcal{S}_i=\mathcal{S}$ such that $\sum_{i=1}^{k}f_i(\mathcal{S}_i)$ is maximized, where the $f_i$'s are arbitrary submodular functions.

\paragraph*{Multiway Greedy Partition Algorithm} To solve the multiway-partitioning problem, a greedy algorithm can be used, which allocates each element $e$ to the best subset at that point, i.e., $\mathcal{S}_{j^{\ast}}\leftarrow \mathcal{S}_{j^{\ast}}\bigcup\{e\}$ where $j^{\ast}=\mathop{\arg\max }_{j} \,f_j(\mathcal{S}_j\bigcup\{e\})$.
If all of the involved functions in the multiway-partitioning problem are non-negative, monotone, and submodular, then the greedy algorithm provides a $1/2$-approximation guarantee \cite{nemhauser1978analysis}.

In the following, we illustrate this problem via a resource allocation application.

\subsection{Water Filling-Based Resource Allocation}

Consider an orthogonal frequency-division multiple access (OFDMA) communication system with a set of $n$ orthogonal subcarriers, denoted by $\mathcal{C}$. The considered problem is to allocate a disjoint subset of $\mathcal{C}$ to each user so as to maximize the sum-rate criterion \cite{7208844}.
Denoting the set of allocated subcarriers to user $i$ by $\mathcal{A}_i \subseteq \mathcal{C}$, the resource allocation problem can be formulated as the following partitioning problem \cite{7208844}:
\begin{align}\label{eq:Partitioning}
&\mathop {\max }\limits_{\mathcal{A}_1,\mathcal{A}_2,\ldots,\mathcal{A}_m} \sum\limits_{i=1}^{m} R_i(\mathcal{A}_i), \nonumber\\
&\text{s.t.}\,\,\, \bigcup_{i=1}^{m} \mathcal{A}_i = \mathcal{C},\, \mathcal{A}_i \cap \mathcal{A}_j=\emptyset,
\end{align}
where
\begin{align}\label{eq:Rate function}
R_i(\mathcal{A}_i) = &\mathop {\max }\limits_{\scriptstyle \sum\nolimits_{j\in\mathcal{A}_i} P_{i,j} \le P_i \hfill\atop \scriptstyle \quad P_{i,j} \ge 0\hfill} \sum_{j\in\mathcal{A}_i} \log\left(1+\frac{P_{i,j}}{N_{i,j}}\right),
\end{align}
with $P_i$ the sum-power constraint for user $i$, $P_{i,j}$ the power allotted by
user $i$ to subcarrier $j$, and $N_{i,j}$ the corresponding channel noise level. The rate \eqref{eq:Rate function} has a similar form as the waterfilling function \cite{7208844}, and thus the power for each user can be allocated locally via a waterfilling algorithm. Furthermore, it is proved that the waterfilling function is submodular \cite{7208844} (since a sum of submodular functions is a submodular function \cite{fujishige2005submodular}, the objective function in \eqref{eq:Partitioning} is submodular) and thus a greedy algorithm can be employed to efficiently solve the subcarrier allocation problem with a theoretical guarantee \cite{nemhauser1978analysis}.

\begin{figure}
	\centering
	\includegraphics[width=.45\textwidth]{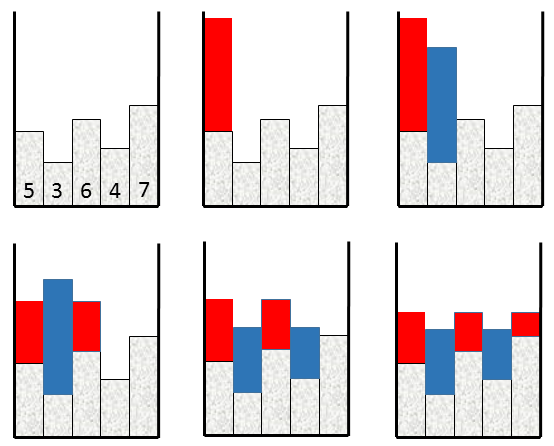}		
	\caption{An example of an uplink OFDMA subcarrier and power allocation problem with two users and five subcarriers. We consider $8$ watt for the sum-power constraint of each user and both users are assumed to experience the same noise variance in each subcarrier. The top left subfigure depicts the noise variance of different subcarriers. Starting from the first channel and allocating the channels one by one, subfigures $2$ to $6$ present the procedure of subcarrier allocation to users, where red/blue bars represent the allocated power by the first/second user to each subcarrier, respectively.}
	\label{fig:Waterfilling_procedure}
\end{figure}

For the presented application, an illustrative example is given in Fig. \ref{fig:Waterfilling_procedure} to demonstrate how a greedy resource allocation algorithm works. As shown in Fig. \ref{fig:Waterfilling_procedure}, starting from the first subcarrier and going through them one by one, at each step, the subcarrier is allocated to the user with the maximum marginal gain and power is reallocated afterward for the new set of subcarriers based on the waterfilling algorithm.

\section{Approximate (Weak) Submodularity}

When the cost function $f$ is not submodular, greedy algorithms can still be useful, yet they do not necessarily provide any theoretical guarantees. Fortunately, for cases where $f$ is \emph{close to} submodular, it can be shown that greedy algorithms degrade gracefully~\cite{das2018approximate}.

To measure how far a function $f$ is from being submodular, the concept of \emph{weak submodularity} has been introduced in~\cite{das2018approximate}. This notion of \emph{approximate} submodularity is linked with other notions such as $\epsilon$-submodularity (see, e.g.,~\cite{krause2010submodular}) and to properties such as the restricted strong convexity \cite{elenberg2018restricted}. Mathematically, weak submodularity is defined through the \emph{submodularity ratio}, $\gamma$, as follows:
\def\cN{\mathcal{N}}
\def\cL{\mathcal{L}}
\def\cS{\mathcal{S}}
A monotone non-negative set function $f : 2^{\mathcal{N}}\rightarrow\mathbb{R}_{+}$ is called $\gamma$-weakly submodular for an integer $r$ if
\begin{equation}\label{eq.subratio}
	\gamma \leq \gamma_r := \underset{\cL,\cS\subseteq\cN:\\\vert\cL\vert,\vert\cS\backslash\cL\vert\leq r}{\min}\frac{\sum_{j\in\cS\backslash\cL}f(\{j\}|\cL)}{f(\cS|\cL)},
\end{equation}
where $0/0$ is defined as $1$.
This definition generalizes submodularity by relaxing the \emph{diminishing returns} property. It can be easily shown that a function $f$ is submodular if and only if $\gamma_{r} \geq 1$ for all $r$. Whenever $\gamma$ is bounded away from $0$, the greedy algorithm guarantees a solution with a $(1-\text{e}^{-\gamma})$--approximation guarantee under a cardinality constraint of size $r$, which is the best achievable performance as shown in \cite{harshaw2019submodular}.
To explore other variants and guarantees for greedy maximization of approximately submodular functions, we refer the readers to~\cite{das2018approximate}.

\subsection{Subset Selection for Regression}

To show how the approximate submodularity framework can be useful in practice, we consider the problem of subset selection for regression \cite{das2018approximate,elenberg2018restricted}. That is, given a set of $n$ regressors, select a subset of $k$ regressors that best predict the variables of interest. The applications of this problem range from feature selection to sparse learning in both signal processing and machine learning. The advantage of using the \emph{natural} combinatorial formulation of the problem, based on weak submodularity, over traditional convex relaxations~\cite{joshi2008sensor} is that it provides a direct control over the sparsity level $k$, and avoids the tuning of regularization parameters.

Consider a set of \emph{observation variables} $\bx = [x_1,\ldots,x_n]^\top$, and a \emph{predictor variable} $z$. Further, let $[\bC]_{i,j} = {\rm Cov}(x_i,x_j)$ and $[\bb]_i={\rm Cov}(x_i,z)$ be the covariances among the observations and between the predictor and observation, respectively. Then, the square multiple correlation with respect to a subset of variables $\cS$ is given by
\begin{equation}
R_{z,\cS}^{2} := \bb_{\cS}^{\top}\bC_{\cS}^{-1}\bb_{\cS},
\label{eq.rz}
\end{equation}
where the subscript indicates either that only the entries or row and column indices in $\cS$ are retained. Hence, given both $\bC$ and $\bb$, and $k$, the subset selection problem is posed as the maximization of $R_{z,\cS}^2 : \vert \cS \vert \leqslant k$. This setting is similar to that of~\cite{coutino2018submodular}, where the approximate submodularity of the signal-to-noise ratio (same functional form as~\eqref{eq.rz}) is leveraged in sensor selection for detection of signals under Gaussian noise.

Using the approximate submodularity, it can be shown that celebrated greedy algorithms such as forward regression (FR) and orthogonal matching pursuit (OMP) obtain near-optimality guarantees for this family of problems. For these two algorithms, their submodularity ratios are given by $\gamma_{\rm FS} = \lambda(4\bC,2k)$ and $\gamma_{\rm OMP} = \lambda(\bC,2k)^2$, respectively. Here, $\lambda(\bC,k) := \min_{\cS : \vert\cS\vert =k}\lambda_{\min}(\bC_{\cS})$, where $\lambda_{\min}(\bA)$ is the minimum eigenvalue of the matrix $\bA$.

\begin{figure*}[h]
	\centering
	\subfigure[] {\includegraphics[width=.45\textwidth]{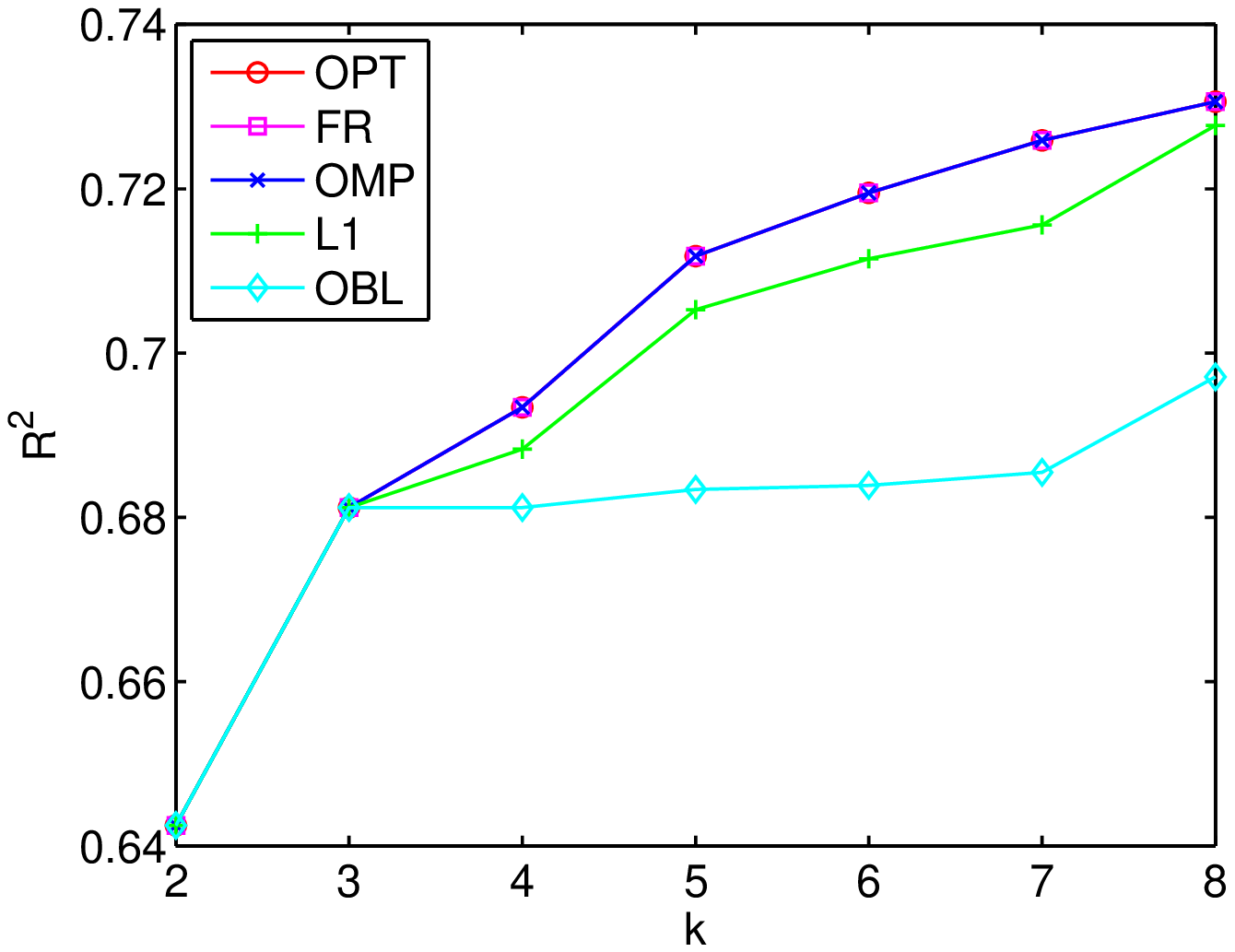}
		}\quad
	\subfigure[] {\includegraphics[width=.45\textwidth]{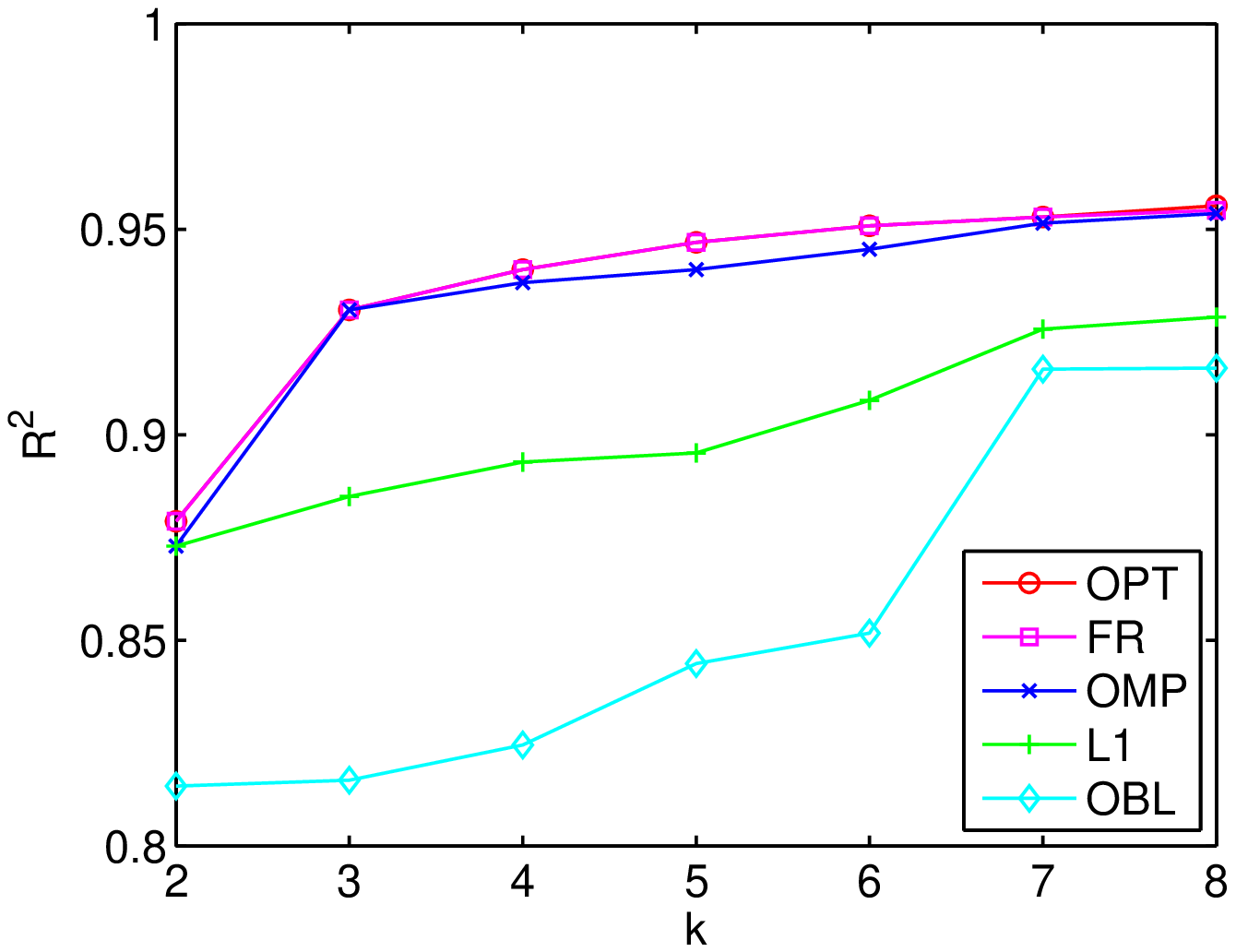}}\\
	\caption{Regression results for two different datasets. (a) Boston Housing dataset   $R^2_{z,\cS}$. (b) World Bank dataset $R^2_{z,\cS}$. Figure courtesy of ~\cite{das2018approximate}}
	\label{fig.regL1}
\end{figure*}

To illustrate this application, a comparison of these two algorithms is conducted against the optimal solution (OPT), the oblivious greedy algorithm (OBL), and the Lasso (L1) algorithm. The theoretical results [cf. the submodularity ratios], predict that FS should outperform OMP in most of the cases. In Fig.~\ref{fig.regL1}, the $R_{z,\cS}^2$ values for the selected subsets by different methods for sizes $k \in \{2,\ldots,8\}$ is shown for two different data sets. 

\section{(Weakly) Adaptive Submodularity}

Several problems in SP and ML require taking decisions in a sequence such as terrain exploration or movie recommendation based on user's feedback. The difficulty of such problems lie not only in the size of the search space but also in the partial knowledge of the process, i.e., decisions are taken under uncertainty of the future. As tackling these problems without considering any further structure is either challenging or intractable, the concept of submodularity has been extended to this setting to find tractable methods with strong theoretical guarantees~\cite{golovin2011adaptive}.

To model adaptability, we can make use of a (directed) graph $G=(\mathcal{V},\mathcal{E})$ whose vertices relate to the elements of the ground set, i.e., there is a bijection between the vertices of $G$ and the elements of the set $\mathcal{N}$; and its weighted (directed) edges capture the probabilities of selecting one of the elements after the other (if this information is available). Adaptability arises when nodes (or edges) are allowed to have \emph{states}. That is, their properties, e.g., cost, signals on the nodes (edges), etc., can change as a selection procedure (creation of a subset of elements) progresses \cite{mitrovic2019adaptive}. 

For instance, let us consider recommending websites to a user. Here the set $\mathcal{N}$ represents websites and the graph $G$ (and its weights) can be constructed based on the traffic (cross-references) between these websites. Further, each website is assumed to have, at any given time, one of the states $\mathcal{Q}:=\{\text{unvisited},\text{visited}\}$. Finally, a function $h:2^{|\mathcal{V}|}\times\mathcal{Q}^{|\mathcal{V}|}\mapsto~ \mathbb{R}_+$ is selected to measure user satisfaction, e.g., affinity of visited websites with dynamic user preferences. Here, adaptability is required as at each new recommendation the states of the websites change.

To extend submodularity notions to the adaptive setting, the conditional marginal gain of a set $\mathcal{A}\subseteq\mathcal{N}$, with respect to the function $h$, is defined as 

\def\dom{{\rm dom}}
\begin{equation}
\Delta(\mathcal{A}|\psi) = \mathbb{E}[h(\dom(\psi)\cup \mathcal{A},\phi)-h(\dom(\psi),\phi)|\psi],
\end{equation}
where $\psi$ denotes a partial realization, i.e., a mapping disclosing the states of a subset of nodes (edges), and the expectation is taken over all the full realizations $\phi$, i.e., complete disclosure of node (edges) states, such that $\dom(\psi)\subseteq\dom(\phi)$. Here $\dom(\psi)$ denotes the list of items whose state is known. Using the above definition, a set function $h$ is \emph{weakly adaptive set submodular} with parameter $\gamma$ if for all sets $\mathcal{A}\subseteq \mathcal{V}$ we have
\begin{equation}
\gamma \leq \frac{\sum_{e\in \mathcal{A}}\Delta(e|\psi)}{\Delta(\mathcal{A}|\psi')}\; \forall\,\psi \subseteq\psi',
\end{equation}
where $\psi \subseteq\psi'$ iff $\dom(\psi)\subseteq\dom(\psi')$ and they are equal in the domain of $\psi$. 

This notion is a natural extension of the submodularity ratio [cf.~\eqref{eq.subratio}] to the adaptive setting. Notice that instead of only conditioning on sets, conditioning on realizations is needed to account for element states. Note that an edge function can be defined in a similar way to $h$.
%}

Under this setting, it has been shown that an adaptive version of the greedy algorithm~\cite{mitrovic2019adaptive}, the adaptive sequence greedy (ASG) method, returns a set $\sigma_{\rm ASG}$, such that
\begin{equation}
f_{\rm avg}(\sigma_{\rm ASG}) \geq \frac{\gamma}{2d_{\rm in}+\gamma}f_{\rm avg}(\sigma^*),
\end{equation}
where $f_{\rm avg}(\sigma):=\mathbb{E}[h(\mathcal{E}(\sigma),\phi)]$; $d_{\rm in}$ is the largest in-degree of the input graph $G$; and $\sigma^*$ is the set obtaining the highest expected value [c.f. $f_{\rm avg}(\cdot)$].

\subsection{Wikipedia Link Search}

To illustrate an application of adaptive sequence submodularity, we present the problem of adaptive article sequence recommendation. Here, a user is surfing Wikipedia towards some target article. And given her history of previously visited links, we aim to guide her to the target article. As the order in which she visits the articles is critical, a set of articles does not suffice as an answer, and an \emph{ordered sequence} is needed.

Under this setting, the sequence value is encoded through the weights of a directed graph $G=(\mathcal{V},\mathcal{E})$, where each element of the ground set is represented by a vertex in $\mathcal{V}$. The probability of moving from the $i$th link to the $j$th link is captured by the weight $w_{ij}$ of the corresponding directed edge. As a result, a sequence of elements $\sigma:=\{\sigma_1,\ldots,\sigma_i,\sigma_{i+1},\ldots\}$, i.e., an ordered set of elements, induces a set of edges $\mathcal{E}(\sigma):=\{(\sigma_i,\sigma_j)|(\sigma_i,\sigma_j)\in \mathcal{E}, i\leq j\}$.  In addition, each node is assumed to have two states: $1$ if the user visits a page and $0$ if the user does not want to visit it. This last feature, plus the fact that the decision must be made based on the current page that the user is visiting, is what makes adaptability necessary. For this problem, the probabilistic coverage utility function~\cite{mitrovic2019adaptive}, i.e.,
\begin{equation}
h(\mathcal{E}(\sigma)) = \sum_{j\in \mathcal{V}}[1 - \prod_{(i,j)\in \mathcal{E}(\sigma)}(1-w_{ij})],
\end{equation}
is used to guide the selection problem.

\begin{figure*}[h]
	\begin{multicols}{2}		
	\centering		
	{\includegraphics[width=1.2\linewidth]{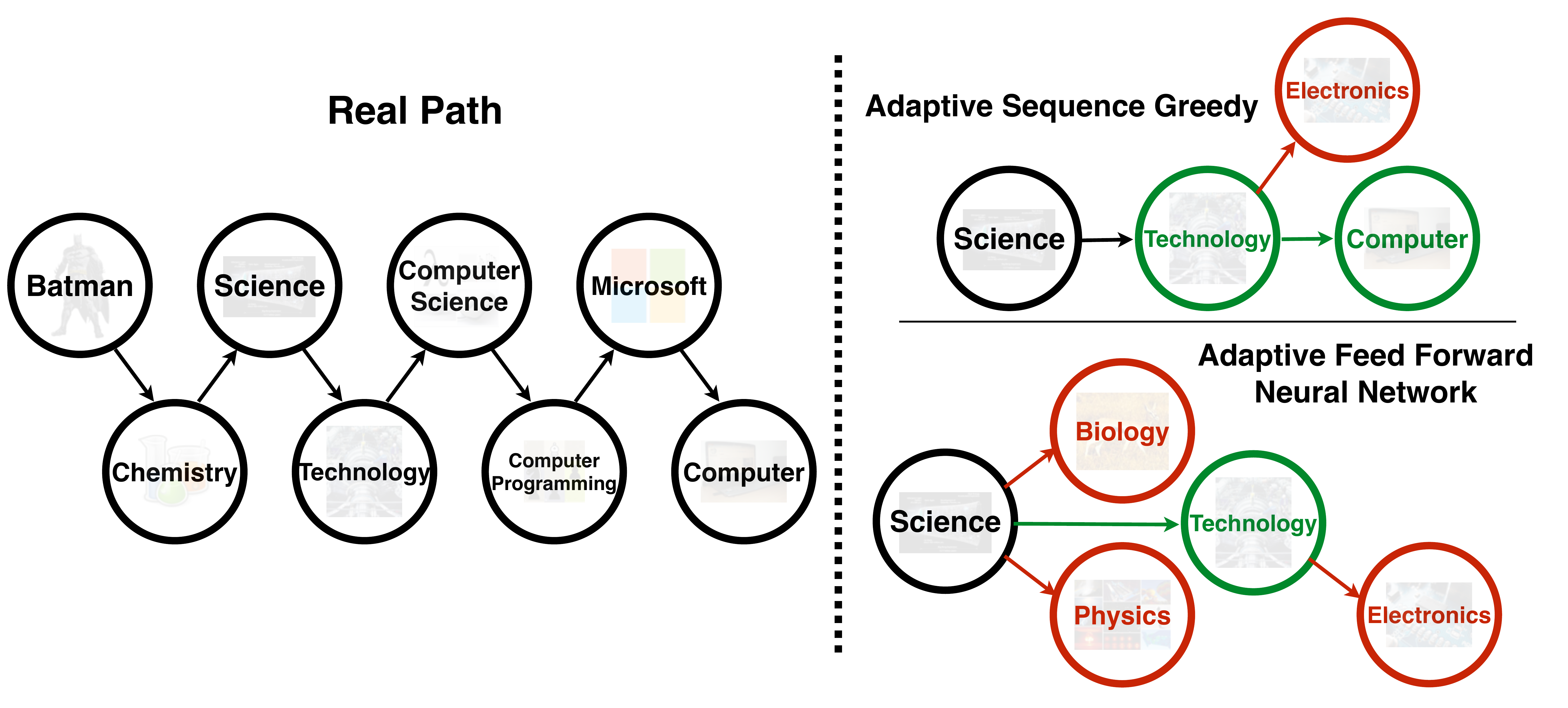}}
	{\includegraphics[width=.7\linewidth]{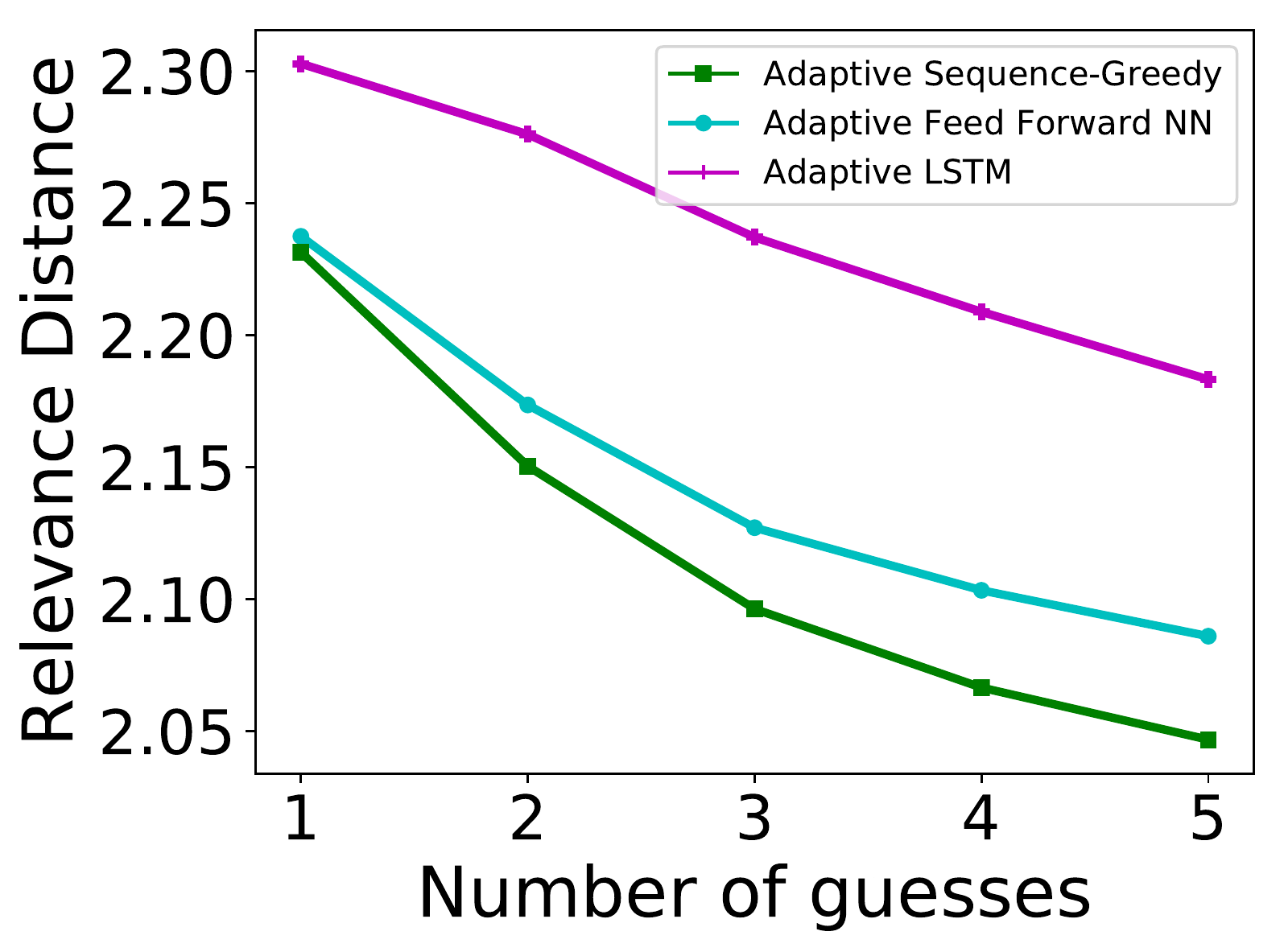}}
	\end{multicols}
	\caption{(Left) Real and (Center) predicted page paths. (Right) Overall performance of the compared methods using the relevance distance. Figure courtesy of~\cite{mitrovic2019adaptive}.}
	\label{fig.dlcomp}
\end{figure*}

A comparison of the ASG method with deep learning based alternatives is shown in Fig.~\ref{fig.dlcomp}. The results report the relevance of the final output page to the true target page, i.e., a higher relevance is related to a lower score. Notice that the ASG method outperforms the deep-learning alternatives as under this setting, we suffer from data scarcity. Also, ASG comes with provable guarantees on its performance and does not require hyperparameter tuning nor retraining. These are in contrast with deep learning approaches which do not have theoretical guarantees, require parameter tuning and, when the ground set is changed, they need to be retrained.

\section{Distributed Submodular Maximization}
As explained so far, although submodularity enables us to employ conceptually low complexity algorithms with theoretical approximation bounds, classical approaches of submodular optimization require access to the full dataset which is impractical in large-scale problems. MapReduce is a fruitful programming model for reliable and efficient parallel processing which has been shown a promising approach in order to design parallel submodular optimization algorithms particularly to form a small representative subset from a large dataset \cite{mirzasoleiman2016distributed, kumar2015fast}.

In a distributed setting, we assume $m$ machines are given to carry out the submodular optimization problem with two considerations: $(i)$ the optimality of the returned solution, and $(ii)$ communication complexity, i.e., the number of synchronizations among machines. One can imagine that without any constraint on the number of synchronizations, we can technically perform a centralized scenario. Consequently, the following three questions arise to tackle distributed optimization: $(i)$ how to distribute items among machines, $(ii)$ what algorithms to run across different machines in a parallel fashion, and $(iii)$ how to merge/synchronize the results of different machines \cite{mirzasoleiman2016distributed}.

\begin{figure}
	\centering
	\includegraphics[width=0.48\textwidth]{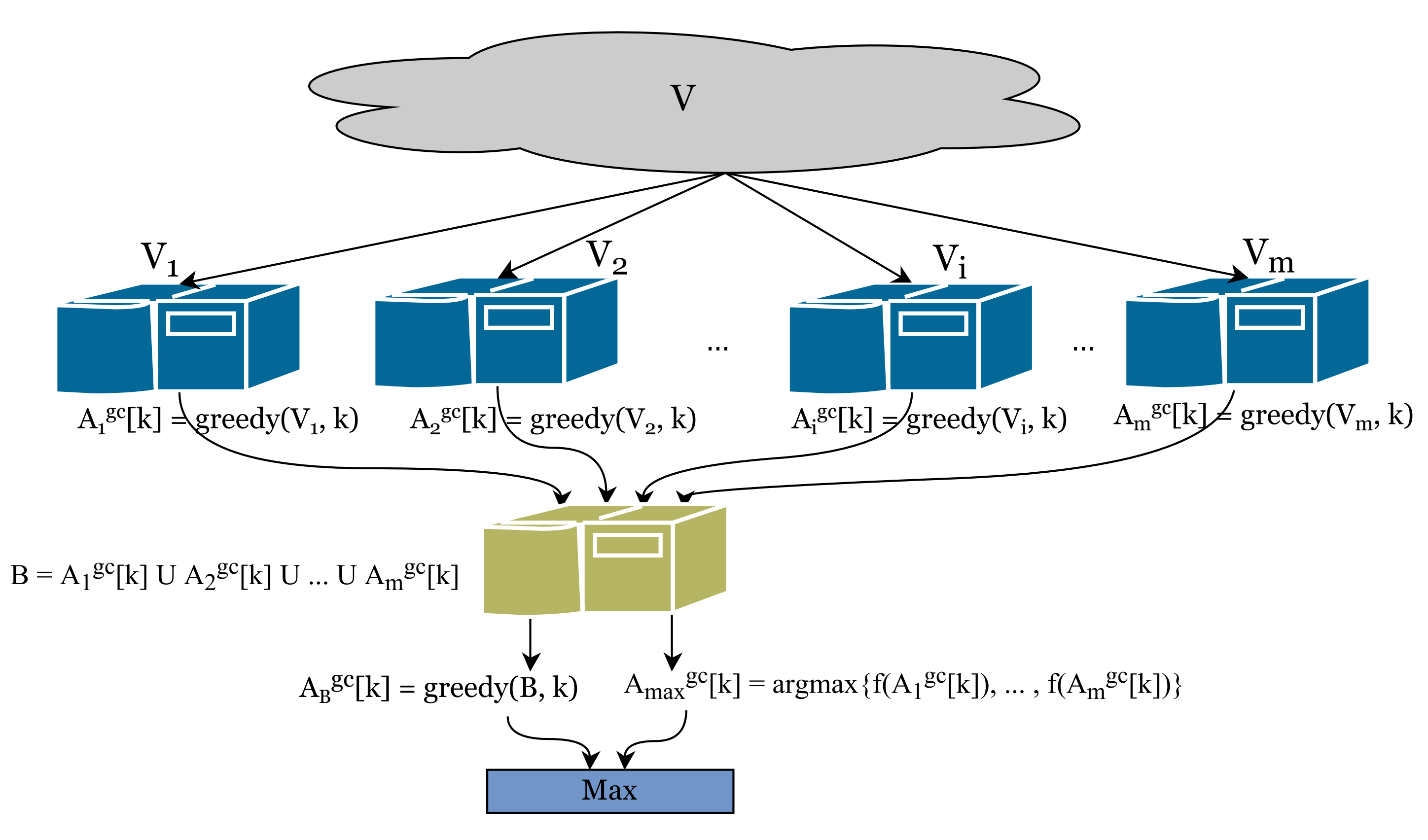}
	\caption{Illustration of the distributed two-round algorithm. $V$ is the ground set and $V_1, \cdots, V_m$ are the initial partitions distributed among different machines with $m$ being the number of machines. Moreover, $A_1^{\rm{gc}}, \cdots, A_m^{\rm{gc}}$ are the greedy solutions obtained locally on machines in the first round where the one with the maximum utility is denoted as $A_{\rm{\max}}^{\rm{gc}}$. Then, in the second round, $m$ subsets obtained in all machines are shared with a central node to form a super set $B$ with $mK$ elements, i.e., $B = A_1^{\rm{gc}} \cup \cdots \cup A_m^{\rm{gc}}$. Furthermore, running a standard greedy algorithm over the super set $B$ leads to $A_B^{\rm{gc}}$. Finally, the maximum of $A_{\rm{\max}}^{\rm{gc}}$ and $A_B^{\rm{gc}}$ is returned as the solution. Figure courtesy of~\cite{mirzasoleiman2016distributed}.}
	\label{fig.DistSub}
\end{figure}

A two-round parallel protocol which provides efficient responses to these three questions is proposed in \cite{mirzasoleiman2016distributed}. Here, we briefly explain the algorithm and result for a monotone submodular function with a cardinality constraint (For more general cases, see \cite{mirzasoleiman2016distributed}).
In the initialization phase, the dataset is arbitrarily partitioned into $m$ sets, one set for each machine. Next, in the first round, given $K$ as the cardinality constraint, each machine executes a greedy algorithm over its own set to achieve a subset with $K$ elements. Then, in the second round, $m$ subsets obtained in all machines are shared with a central node to form a super set with $mK$ elements. Running a standard greedy algorithm over this super set leads to a new subset with $K$ elements. Finally, among the $m+1$ subsets with $K$ elements, the one which maximizes the utility function is selected. It is shown that this algorithm provides a $\frac{(1-1/{\rm{e}})}{\sqrt{\min(m,K)}}$-approximation guarantee \cite{mirzasoleiman2015distributed}. Fig. \ref{fig.DistSub} depicts an illustration of the two-round algorithm in \cite{mirzasoleiman2016distributed}. Active set selection in Gaussian processes and large-scale exemplar based clustering are two instances of applications where the size of the datasets often requires a distributed method for a given submodular maximization problem \cite{mirzasoleiman2016distributed}.

\section{Submodularity and Convexity: Minimization}

Up to this point, we have only discussed maximization of submodular functions. In this section, we highlight an interesting connection between submodular set functions and convexity that allows the efficient minimization of submodular set functions.

\subsection{A Convex Extension of Submodular Functions}
Consider a set function $f:2^{\mathcal{N}}\rightarrow \mathbb{R}$ with $|\mathcal{N}|=N$. Associate every element of the set $2^{\mathcal{N}}$ to a vertex of the hypercube $\{0,1\}^N$. That is, each $\mathcal{S}\subseteq \mathcal{N}$ corresponds uniquely to a binary vector of length $N$ where the $n$th entry is $1$ if $n\in\mathcal{S}$ and $0$ otherwise.

\begin{figure*}
	\centering
	\includegraphics[width=.9\textwidth]{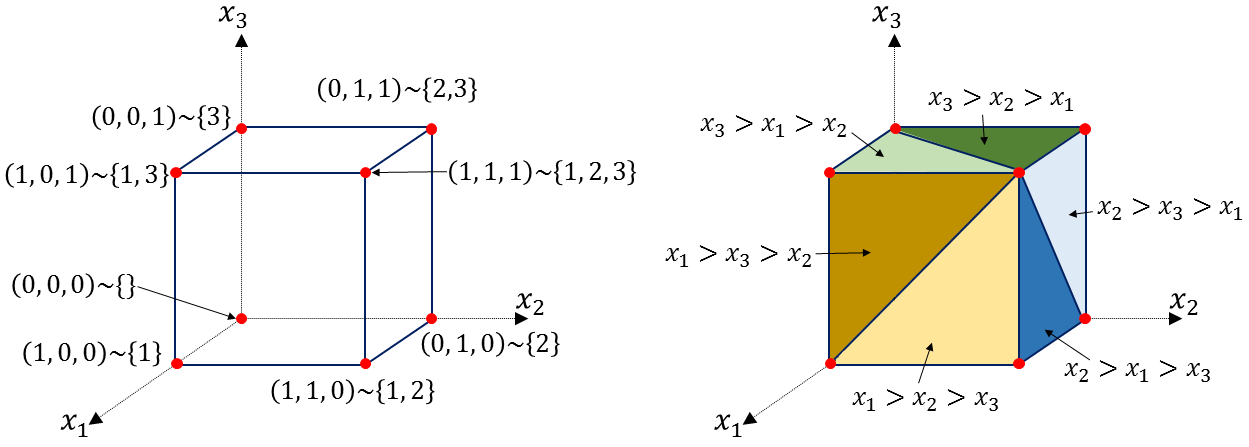}		
	\caption{From subsets to vertices in the hypercube for a sample ground set $\mathcal{N}=\{1,2,3\}$ with $N=3$. Each subset has a corresponding vertex in the hypercube. The left subfigure depicts how the hypercube is created. The right subfigure presents the $6$ possible orderings of the input vector for the Lov{\'a}sz extension~\cite{bach2013learning}.}
	\label{Lovasz_illustrate}
\end{figure*}

A set function $f$ can be extended from the discrete domain (vertices of the hypercube) to the continuous domain (the complete hypercube) through the Lov{\'a}sz extension. The Lov{\'a}sz extension $F_L:\mathbb{R}^N \rightarrow \mathbb{R}$, is defined as follows: 
given a vector $\bx \in [0,1]^N$ and a scalar $\theta\in[0,1]$, define $\mathcal{T}_{\theta}=\{e\in\mathcal{N}|x_e\geqslant \theta\}$, where $x_e$ is the $e$th entry of $\bx$. Then, $F_L(\bx)$ is obtained through the following equation \cite{lovasz1983submodular}:

\begin{equation}
F_L(\bx) = \underset{\theta\in[0,1]}{\mathbb{E}}\{f(\mathcal{T}_{\theta})\}.
\end{equation}
Fig. \ref{Lovasz_illustrate} illustrates an example of how the hypercube is formed and is divided into six parts corresponding to the six possible orderings of the input vector for the Lov{\'a}sz extension.

The Lov{\'a}sz extension of a submodular function has two main properties \cite{lovasz1983submodular}: 
First, it is convex; and second, its minimizer resides at the vertices of the hypercube. Hence, convex optimization can be employed to find efficiently a feasible minimizer of the original discrete problem.

We note that, in both Lov{\'a}sz  and multi-linear extensions, the distribution defining the continuous function at $\bx$ is independent of the set function $f(\cdot)$. Also, these extensions are similar in the sense that both are obtained by taking the expectation of the function however with respect to different probability measures.

\subsection{Unconstrained Submodular Minimization}
Consider an unconstrained submodular minimization problem, i.e., $\min_{\mathcal{S}\subseteq\mathcal{N}}f(\mathcal{S})$. If $f$ is extended to $F_{L}$, the continuous function is \emph{convex} and \emph{exact} (i.e., we can recover an optimal set of the original problem from an optimal solution of the extended problem). Therefore, solving $\min_{\bx\in[0,1]^N} F_L(\bx)$ by a subgradient method, we can retrieve the optimal solution for the original discrete problem. 

Alternatively, using the dual formulation of the continuous problem, we can obtain a minimum norm problem which can be solved by the Frank-Wolfe algorithm \cite{frank1956algorithm}. This algorithm, also known as the conditional gradient method, is an iterative first-order method that considers a linear approximation of the objective function 
and moves towards its minimizer. The Frank-Wolfe algorithm is a projection-free method and it is well-known for keeping the sparsity of the candidate solution. These aspects make this algorithm attractive for sparse and large scale constrained optimization problems, for instance, in optimizing over atomic domains \cite{jaggi2013revisiting}.

\subsection{Hardness of Constrained Submodular Minimization}
Although there exist polynomial-time algorithms for minimizing any unconstrained submodular function, constrained submodular minimization becomes challenging to approximate under simple constraints. For instance, in \cite{svitkina2011submodular}, it is shown that for problems such as submodular load balancing (given a monotone submodular function $f$ and a positive integer $M$, find a partition of $\mathcal{N}$ into $M$ sets, $\mathcal{N}_1, \cdots , \mathcal{N}_M$, so as to minimize $\underset{m}{\max}~ f(\mathcal{N}_m)$) and 
Submodular Sparsest Cut (given a set of unordered pairs $\{\{x_m, y_m\} | x_m, y_m \in \mathcal{N} \}$, each with a demand $p_m > 0$, find a subset $\mathcal{S} \subseteq \mathcal{N}$ minimizing $\frac{f(\mathcal{S})}{\sum_{m:|\mathcal{S}\cap\{x_m,y_m\}|=1}p_m}$), %and submodular function minimization with a cardinality lower bound, 
the approximation guarantees cannot be better than $O(\sqrt{\frac{N}{\ln N}})$ where $N$ is the size of the ground set.

Therefore, even though polynomial-time algorithms are available for the minimization of submodular functions, we must be aware of the following issues: $(i)$ high computational complexity of algorithms to minimize unconstrained problems, i.e., the polynomial orders are typically larger than $3$, making exact minimization challenging for large-scale problems; and $(ii)$ dealing with constraints makes the problem extremely hard. %As explained next, these issues do not appear in the case of submodular maximization, even though it is NP-hard in general.
%}

\section{Optimization of General Set Functions}

At this point, readers may wonder: Is it possible to develop a suitable greedy algorithm for any arbitrary set function? Unfortunately, the answer is no, in general. 
However, using the fact that any set function can be expressed as a difference of two submodular set functions \cite{narasimhan2012submodular}, greedy algorithms inspired by optimization methods for the difference of two convex functions~\cite{yuille2003concave} can be developed.

The maximization of any set function $f: 2^{\mathcal{N}} \rightarrow \mathbb{R}$, defined over a ground set $\mathcal{N}$, can be expressed as the difference of two submodular set functions $g: 2^{\mathcal{N}} \rightarrow \mathbb{R}$ and $h: 2^{\mathcal{N}} \rightarrow \mathbb{R}$:
\begin{align}\label{eq:Problem SupSub}
\mathop {\max }\limits_{\mathcal{S} \subseteq \mathcal{N}} f\left( \mathcal{S} \right) \equiv \mathop {\max }\limits_{\mathcal{S} \subseteq \mathcal{N}} \left[ {g\left( \mathcal{S} \right) - h\left( \mathcal{S} \right)} \right].
\end{align}

This formulation allows for drawing parallels with convex optimization techniques and to devise a greedy algorithm to approximate the solution.
Specifically when $h(\mathcal{S})$ is modular, a recent result shows that we can get a $(1-1/\text{e})g(\text{opt})-h(\text{opt})$ approximation guarantee \cite{harshaw2019submodular}. 

\subsection{SupSub Procedure}

Similar to maximizing the difference of convex functions, we can consider an approach that approximates the solution of the original problem by a sequence of submodular maximization problems. Recall that in the \emph{convex-concave} procedure~\cite{yuille2003concave}, the concave function is approximated at every step by its first-order Taylor expansion.

Following this idea, the problem of maximizing the difference of submodular set functions is cast as the sequential maximization of submodular functions. This is done by substituting the second submodular set function in \eqref{eq:Problem SupSub} with its modular upper bound in each iteration \cite{iyer2012algorithms}. A number of tight modular upper bounds $\mathcal{J}_{\mathcal{S}}^h$ are suggested in \cite{nemhauser1978analysis}, e.g.,
\begin{align}
\mathcal{J}_{\mathcal{S}}^h(\mathcal{X}) & = h(\mathcal{S}) - \sum\limits_{e \,\in\, \mathcal{S}\backslash \mathcal{X}} {h\left( {\{ e\} |\mathcal{S}\backslash \{ e\} } \right)}  + \sum\limits_{e \,\in\, \mathcal{X}\backslash \mathcal{S}} {h\left( {\{ e\} |\emptyset } \right)}.
\end{align}

This method, called \textit{supermodular-submodular} (SupSub) procedure, starts with an empty set $\mathcal{S}_0$ and in each iteration $k$ tries to solve the following problem (specialized for a $K$-cardinality constraint):
\begin{align}
\mathcal{S}_{k+1}=\mathop{\arg\max } \limits_{\mathcal{S} \subseteq \mathcal{N}, \,|\mathcal{S}|=K} \,g(\mathcal{S})-\mathcal{J}^h_{\mathcal{S}_k}(\mathcal{S})
\end{align}
until a convergence condition is satisfied. 

Notice that to solve the maximization problem at each step of this algorithm, which is NP-hard in general, the greedy heuristic can be used to obtain a near-optimal solution.	Despite that near-optimality guarantees are not available for the SupSub procedure, similar to the convex-concave procedure, this method is guaranteed to converge to a local minima~\cite{iyer2012algorithms}.

\subsection{Feature Selection for Classification}

We consider the binary classification problem as a non-submodular example to which the SupSub procedure can be applied.
Note that exploiting more features does not necessarily reduce the classification error, however, we should find a way to select the most informative features for a given dataset as quickly as possible.
Thus, a greedy feature selection procedure can be considered as an attractive solution.

Consider the binary classification problem, modeled as a binary hypothesis test given by
\begin{equation}\label{eq:Hypotheis general}
\begin{cases}
&\mathcal{H}_0:\,\mathbf{y}_{\mathcal{S}} \sim \mathcal{N}(\boldsymbol{\theta}_{0,\mathcal{S}},\mathbf{\Sigma}_{0,\mathcal{S}}) \\
&\mathcal{H}_1:\,\mathbf{y}_{\mathcal{S}} \sim \mathcal{N}(\boldsymbol{\theta}_{1,\mathcal{S}},\mathbf{\Sigma}_{1,\mathcal{S}})
\end{cases}
\end{equation}
where $\mathcal{S}\subseteq \mathcal{N}$ is the subset of selected features from the ground set $\mathcal{N}=\{1,\ldots,N\}$.
The mean vectors of the selected data under $\mathcal{H}_0$ and $\mathcal{H}_1$ are denoted by $\boldsymbol{\theta}_{0,\mathcal{S}}$ and $\boldsymbol{\theta}_{1,\mathcal{S}}$, and the second-order statistics by $\mathbf{\Sigma}_{0,\mathcal{S}}$ and $\mathbf{\Sigma}_{1, \mathcal{S}}$, respectively.

We consider the Kullback-Leibler (KL) divergence, $\mathcal{K}(\mathcal{H}_1||\mathcal{H}_0)$, as the performance measure for the classification task, which is a distance measuring how far the two hypotheses $\mathcal{H}_0$ and $\mathcal{H}_1$ are.
Unfortunately, the KL divergence is not submodular. To solve this problem, the KL divergence was decomposed in \cite{coutino2018submodular} as a difference of two submodular set functions, which means the SupSub procedure can be employed.

\begin{figure}[htb]
	%\vspace{-.3cm}
	\centering
	\centerline{\includegraphics[width=0.5\textwidth]{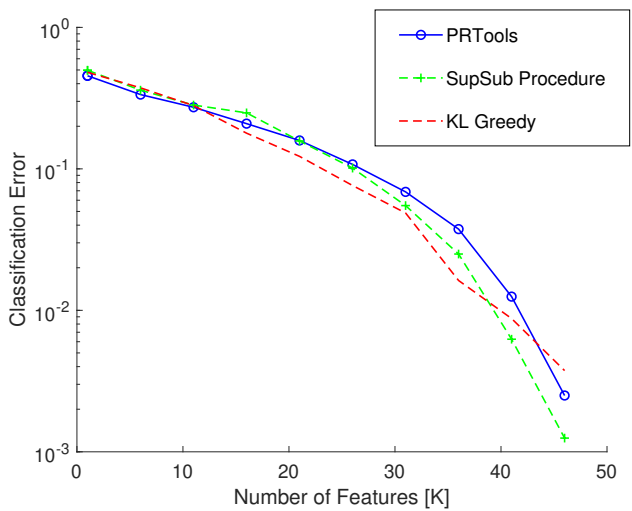}}
	%\vspace{-.2cm}
	\caption{ Classification soft error when using QDC for a Gaussian binary classification problem. Figure courtesy of~\cite{coutino2018submodular}. }
	\label{fig:Feature Selection1}
	%\vspace{-.2cm}
\end{figure}

Now, let us assess the performance of the greedy method for feature selection. In this example, we consider two classes described by a Gaussian distribution with second-order statistics $\mathbf{\Sigma}_0$ and $\mathbf{\Sigma}_1$, and mean vectors $\boldsymbol{\theta}_0$ and $\boldsymbol{\theta}_1$, respectively. Here, the covariance matrices are assumed to be Toeplitz matrices (common structure in signal processing problems). The total number of features is 50, and the trained classifier is the quadratic discriminant classifier (QDC).
In Fig.~\ref{fig:Feature Selection1}, the classification soft error for different feature selection methods is depicted versus the cardinality of the selected feature set. The feature selection method in the PRTools Toolbox is considered here as a baseline \cite{coutino2018submodular}.
As shown in this figure, the method based on the SupSub procedure provides a desirable performance superior to the PRTools baseline result. Furthermore, the KL greedy, i.e., directly applying the greedy heuristic to the KL function, outperforms the other methods in most cases. However, it can get stuck sometimes and has no near-optimality guarantees (for further discussions, see \cite{coutino2018submodular}).

\section{Submodularity and Continuous Domain Optimization}
As discussed in previous sections, submodularity is a useful property of functions defined in a discrete domain which admits a guaranteed approximate optimization with efficient algorithms.
The readers may expect that an extension of submodularity to the continuous domain provides similar benefits for continuous optimization problems.

In this regard, in \cite{bach2019submodular}, the notion of continuous submodularity is defined on subsets of $\mathbb{R}^N$ so that a function $f:\mathcal{X}\rightarrow\mathbb{R}$, where $\mathcal{X}=\prod_{i=1}^N{\mathcal{X}_i}$ with each $\mathcal{X}_i$ is an interval, is continuous submodular \textit{iff} for all $(\mathbf{x},\mathbf{y})\in \mathcal{X}\times \mathcal{X}$, 
\begin{align}\label{eq:Def. cont. Submod.}
f(\mathbf{x})+f(\mathbf{y})\ge f(\mathbf{x}\vee\mathbf{y}) + f(\mathbf{x}\wedge\mathbf{y})
\end{align}
where $\vee$ and $\wedge$ stand for the coordinate-wise maximum and minimum operators, respectively. When $f(\cdot)$ is twice-differentiable, this function is submodular \textit{iff} all non-diagonal elements of its Hessian are non-positive \cite{bach2019submodular}.

The class of continuous submodular functions covers a subset of both convex and concave functions. As an example, a function of the form $f_{i,j}(x_i-x_j)$ for a convex $f_{i,j}$ is both submodular and convex; or an indefinite quadratic function of the form $f(\mathbf{x})=\frac{1}{2}\mathbf{x}^T\mathbf{A}\mathbf{x}+\mathbf{b}^T\mathbf{x}+c$
with all non-diagonal elements of $\mathbf{A}$ non-positive is a submodular but non-convex/non-concave function.

Analogues to the discrete domain, the diminishing returns (DR) property is generalized to functions defined over $\mathcal{X}$ (see \cite{bian2016guaranteed}).
It is clear that for set functions, the DR property is equivalent to submodularity, however, for general continuous domain functions, submodularity does not necessarily imply the DR property.
In other words, the DR property is stronger than submodularity in general.
If a continuous submodular function is coordinate-wise concave, it satisfies the DR property \cite{bian2016guaranteed}, which defines a subclass of submodular functions called DR-submodular.
Being twice-differentiable, DR-submodularity is equivalent to the non-positivity of all Hessian entries.

One can exploit the well-known gradient ascent algorithm to maximize a continuous submodular function, which achieves a 1/2-approximation guarantee \cite{hassani2017gradient}.
To gain a superior guarantee, in \cite{bian2016guaranteed}, a variant of the Frank-Wolfe algorithm for maximizing a monotone DR-submodular continuous function under down-closed convex constraints has been proposed, providing a $(1-1/e)$-approximation guarantee. Recently in \cite{mokhtari2017conditional}, a stochastic continuous greedy algorithm has been developed, achieving a $(1-1/e)$-approximation guarantee, which deals with maximizing a similar optimization problem subject to a general convex body constraint. 
Maximizing non-monotone continuous DR-submodular functions has also been studied in \cite{bian2016guaranteed,bian2017continuous, mokhtari2018stochastic}.
Furthermore, the problem of submodular continuous function minimization has been considered in \cite{bach2019submodular}, which proved that efficient techniques from convex optimization can be employed for this task.

\subsection{Non-Convex/Non-Concave Quadratic Function Maximization}
\begin{figure}
	\centering
	\includegraphics[width=0.45\textwidth]{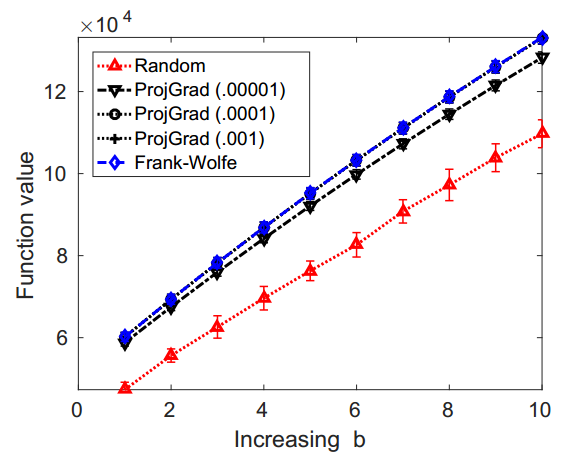}
	\caption{The objective function value as a function of different values of the upper-bound $b$. Figure courtesy of \cite{bian2016guaranteed}.}
	\label{fig.NQP}
\end{figure}

Non-convex/non-concave quadratic programming under general convex constraints arises in various applications, including price optimization, scheduling, graph theory, and free boundary problems, to name a few.
A special class of such problems is submodular quadratic programming which can be tractably optimized.
In this example, a monotone DR-submodular quadratic program is generated under the positive polytope constraint $\mathcal{P}=\{\mathbf{x}\in\mathbb{R}^N|\mathbf{Ax}\le \mathbf{b}, \mathbf{0}\le\mathbf{x}\le\mathbf{1}\}$, where $\mathbf{A}$ has uniformly distributed entries in the interval $[0,1]$, $\mathbf{b}=b\mathbf{1}$ and $N=100$.
In Fig.~\ref{fig.NQP}, the value of the objective function obtained by the Frank-Wolfe variant \cite{bian2016guaranteed} is compared, as a function of $b$, with that of the random and empirically tuned projected gradient method \cite{hassani2017gradient} for three different step sizes. It is noteworthy that the Frank-Wolfe variant provides provable performance guarantees without any tuning requirement, while the performance of the projected gradient is sensitive to parameter tuning.

\section{Conclusion and Research Trends}
In this paper, we explained the concept of submodularity, provided the intuition of how it works, and illustrated some properties. The connection with the convexity in the continuous domain was discussed and the minimization problem was briefly explained.
Also, the concavity aspect of submodularity was demonstrated along with low computational complexity algorithms to maximize submodular functions where the corresponding theorems that guarantee a near-optimal solution were presented.
Moreover, several applications in SP and ML have been covered to transfer the flavor of submodularity to practice. However, it should be pointed out that there is a vast literature on submodularity with a wide variety of applications that were not covered in this paper for the sake of conciseness. 
Continuous submodularity is one of the ongoing research directions that finds applications in robust resource allocation \cite{staib2017robust}.
Online submodular optimization is another research trend that opens up opportunities for many applications such as experimental design \cite{chen2018online}. Finally, it is worth mentioning that submodular optimization is an active research area that is growing fast not only through proposing new algorithms and theories but also via introducing new applications.

\ifCLASSOPTIONcaptionsoff
\newpage
\fi
\bibliographystyle{ieeetr}
\bibliography{ref}

\end{document}